# The diminishing role of hubs in dynamical processes on complex networks

Rick Quax, Andrea Apolloni and Peter M. A. Sloot

| | |
|---|---|
| **Supplementary data** | "Data Supplement"<br>http://rsif.royalsocietypublishing.org/content/suppl/2013/08/29/rsif.2013.0568.DC1.html |
| **References** | **This article cites 66 articles, 9 of which can be accessed free**<br>http://rsif.royalsocietypublishing.org/content/10/88/20130568.full.html#ref-list-1 |
| 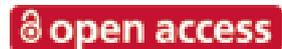 | This article is free to access |
| **Email alerting service** | Receive free email alerts when new articles cite this article - sign up in the box at the top right-hand corner of the article or click **here** |

**Author for correspondence:**
Rick Quax
e-mail: r.quax@uva.nl

†Present address: London School of Hygiene and Tropical Medicine, Keppel Street, London WC1E 7HT, UK.

# The diminishing role of hubs in dynamical processes on complex networks

Rick Quax[1], Andrea Apolloni[2,†] and Peter M. A. Sloot[1,3,4]

[1]Computational Science, University of Amsterdam, Science Park 904, 1098 XH Amsterdam, The Netherlands
[2]Institut des Systèmes Complexes Rhone-Alpes (IXXI) and Laboratoire de Physique, École Normale Supérieure de Lyon, 46 Allée d'Italie, 69007 Lyon, France
[3]National Research University of Information Technologies, Mechanics and Optics (ITMO), Kronverkskiy 49, 197101 Saint Petersburg, Russia
[4]College of Engineering, Nanyang Technological University, 50 Nanyang Avenue, 639798 Singapore, Republic of Singapore

It is notoriously difficult to predict the behaviour of a complex self-organizing system, where the interactions among dynamical units form a heterogeneous topology. Even if the dynamics of each microscopic unit is known, a real understanding of their contributions to the macroscopic system behaviour is still lacking. Here, we develop information-theoretical methods to distinguish the contribution of each individual unit to the collective out-of-equilibrium dynamics. We show that for a system of units connected by a network of interaction potentials with an arbitrary degree distribution, highly connected units have less impact on the system dynamics when compared with intermediately connected units. In an equilibrium setting, the hubs are often found to dictate the long-term behaviour. However, we find both analytically and experimentally that the instantaneous states of these units have a short-lasting effect on the state trajectory of the entire system. We present qualitative evidence of this phenomenon from empirical findings about a social network of product recommendations, a protein–protein interaction network and a neural network, suggesting that it might indeed be a widespread property in nature.

## 1. Introduction

Many non-equilibrium systems consist of dynamical units that interact through a network to produce complex behaviour as a whole. In a wide variety of such systems, each unit has a state that quasi-equilibrates to the distribution of states of the units it interacts with, or 'interaction potential', which results in the new state of the unit. This assumption is also known as the local thermodynamic equilibrium (LTE), originally formulated to describe radiative transfer inside stars [1,2]. Examples of systems of coupled units that have been described in this manner include brain networks [3–6], cellular regulatory networks [7–11], immune networks [12,13], social interaction networks [14–20] and financial trading markets [15,21,22]. A state change of one unit may subsequently cause a neighbour unit to change its state, which may, in turn, cause other units to change, and so on. The core problem of understanding the system's behaviour is that the topology of interactions mixes cause and effect of units in a complex manner, making it hard to tell which units drive the system dynamics.

The main goal of complex systems research is to understand how the dynamics of individual units combine to produce the behaviour of the system as a whole. A common method to dissect the collective behaviour into its individual components is to remove a unit and observe the effect [23–32]. In this manner, it has been shown, for instance, that highly connected units or hubs are crucial for the structural integrity of many real-world systems [28], i.e. removing only a few hubs disconnects the system into subnetworks which can no longer interact. On the other hand, Tanaka et al. [32] find that sparsely connected units are crucial for the dynamical integrity of systems where the remaining (active) units must compensate for the removed (failed) units. Less attention has been paid to study the interplay of the unit dynamics and network topology, from which the system's behaviour emerges, in a non-perturbative and unified manner.

We introduce an information-theoretical approach to quantify to what extent the system's state is actually a representation of an instantaneous state of an individual unit. The minimum number of yes/no questions that is required to determine a unique instance of a system's state is called its *entropy*, measured in the unit *bits* [33]. If a system $S^t$ can be in state $i$ with probability $p_i$, then its Shannon entropy is

$$H(S^t) = -\sum_i p_i \log_2 p_i. \quad (1.1)$$

For example, to determine a unique outcome of $N$ fair coin flips requires $N$ bits of *information*, that is, a reduction of entropy by $N$ bits. The more bits of a system's state $S^t$ are determined by a prior state $s_i^{t_0}$ of a unit $s_i$ at time $t_0$, the more the system state depends on that unit's state. This quantity can be measured using the mutual information between $s_i^{t_0}$ and $S^t$, defined as

$$I(S^t; s_i^{t_0}) = H(S^t) - H(S^t | s_i^{t_0}), \quad (1.2)$$

where $H(X|Y)$ is the conditional variant of $H(X)$. As time passes ($t \to \infty$), $S^t$ becomes more and more independent of $s_i^{t_0}$ until eventually the unit's state provides zero information about $S_t$. This mutual information integrated over time $t$ is a generic measure of the extent that the system state trajectory is dictated by a unit.

We consider large static networks of identical units whose dynamics can be described by the Gibbs measure. The Gibbs measure describes how a unit changes its state subject to the combined potential of its interacting neighbours, in case the LTE is appropriate and using the maximum-entropy principle [34,35] to avoid assuming any additional structure. In fact, in our LTE description, each unit may even be a subsystem in its own right in a multi-scale setting, such as a cell in a tissue or a person in a social network. In this viewpoint, each unit can actually be in a large number of (unobservable) microstates which translate many-to-one to the (observable) macrostates of the unit. We consider that at a small timescale, each unit probabilistically chooses its next state depending on the current state of its neighbours, termed discrete-time Markov networks [36]. Furthermore, we consider random interaction networks with a given degree distribution $p(k)$, which denotes the probability that a randomly selected unit has $k$ interactions with other units, and which have a maximum degree $k_{max}$ that grows less than linear in the network size $N$. Self-loops are not allowed. No additional topological features are imposed, such as degree–degree correlations or community structures. An important consequence of these assumptions for our purpose is that the network is 'locally tree-like' [37,38], i.e. link cycles are exceedingly long.

We show analytically that for this class of systems, the impact of a unit's state on the short-term behaviour of the whole system is a decreasing function of the degree $k$ of the unit for sufficiently high $k$. That is, it takes a relatively short time-period for the information about the instantaneous state of such a high-degree unit to be no longer present in the information stored by the system. A corollary of this finding is that if one would observe the system's state trajectory for a short amount of time, then the (out-of-equilibrium) behaviour of the system cannot be explained by the behaviour of the hubs. In other words, if the task is to optimally predict the short-term system behaviour after observing a subset of the units' states, then high-degree units should not be chosen.

We validate our analytical predictions using numerical experiments of random networks of 6000 ferromagnetic Ising spins where the number of interactions $k$ of a spin is distributed as a power-law $p(k) \propto k^{-\gamma}$. Ising-spin dynamics are extensively studied and are often used as a first approximation of the dynamics of a wide variety of complex physical phenomena [37]. We find further qualitative evidence in the empirical data of the dynamical importance of units as function of their degree in three different domains, namely viral marketing in social networks [39], evolutionary conservation of human proteins [40] and the transmission of a neuron's activity in neural networks [41].

## 2. Results

### 2.1. Information dissipation time of a unit

As a measure of the dynamical importance of a unit $s$, we calculate its information dissipation time (IDT), denoted $D(s)$. In words, it is the time it takes for the information about the state of the unit $s$ to disappear from the network's state. As another way of describing it, it is the time it takes for the network as a whole to forget a particular state of a single unit. Here, we derive analytically a relation between the number of interactions of a unit and the IDT of its state. Our method to calculate the IDT is a measure of cause and effect and not merely of correlation; see appendix for details.

#### 2.1.1. Terminology

A system $S$ consists of units $s_1, s_2, \ldots$ among which some pairs of units, called edges, $E = (s_i, s_j), (s_k, s_l), \ldots$ interact with each other. Each interaction is undirected, and the number of interactions that involve unit $s_i$ is denoted by $k_i$, called the *degree*, which equals $k$ with probability $p(k)$, called the degree distribution. The set of $k_i$ units that $s_i$ interacts with directly is denoted by $h_i = \{x : (s_i, x) \in E\}$. The state of unit $s_i$ at time $t$ is denoted by $s_i^t$, and the collection $S^t = s_1^t, s_2^t, \ldots, s_N^t$ forms the state of the system. Each unit probabilistically chooses its next state based on the current state of each of its nearest-neighbours in the interaction network. Unit $s_i$ chooses the next state $x$ with the conditional probability distribution $p(s_i^{t+1} = x | h_i^t)$. This is also known as a Markov network.

#### 2.1.2. Unit dynamics in the local thermodynamic equilibrium

Before we can proceed to show that $D(s)$ is a decreasing function of the degree $k$ of the unit $s$, we must first define the class of unit dynamics in more detail. That is, we first specify an expression for the conditional probabilities $p(s^{t+1} = r | h^t)$.

We focus on discrete-time Markov networks, so the dynamics of each unit is governed by the same set of conditional probabilities $p(s^{t+1} = r | h^t)$ with the Markov property. In our LTE description, a unit chooses its next state depending on the energy of that state, where the energy landscape induced by the states of its nearest-neighbours through its interactions. That is, each unit can quasi-equilibrate its state to the states of its neighbours. The higher the energy of a state at a given time, the less probable the unit chooses the state. Stochasticity can arise if multiple states have an equal energy, and additional stochasticity is introduced by means of the temperature of the heat bath that surrounds the network.

The consequence of this LTE description that is relevant to our study is that the state transition probability of a unit is an exponential function with respect to the energy. That is, in a discrete-time description, $s^t$ chooses $s^{t+1} = r$ as the next state with a probability

$$p(s^{t+1} = r | h^t) \propto \exp \sum_{s_j \in h} \frac{-e(r | s_j^t)}{T}, \quad (2.1)$$

where $T$ is the temperature of the network's heat bath and $\sum_j e(r | s_j^t)$ is the energy of state $r$ given the states of its interacting neighbours $s_j^t \in h^t$. As a result, the energy landscape of $r$ does not depend on individual states of specific neighbour units; it depends on the distribution of neighbour states.

### 2.1.3. Information as a measure of dynamical impact

The instantaneous state of a system $S^t$ consists of $H(S^t)$ bits of Shannon information. In other words, $H(S^t)$ answers to unique yes/no questions (bits) must be specified in order to determine a unique state $S^t$. As a consequence, the more bits about $S^t$ are determined by the instantaneous state $s_i^{t_0}$ of a unit $s_i$ at time $t_0 \leq t$, the more the system state $S^t$ depends on the unit's state $s_i^{t_0}$.

The impact of a unit's state $s_i^{t_0}$ on the system state $S^t$ at a particular time $t$ can be measured by their mutual information $I(S^t; s_i^{t_0})$. In the extreme case that $s_i^{t_0}$ fully determines the state $S^t$, the entropy of the system state coincides with the entropy of the unit state, and the dynamical impact is maximum at $H(S^t) = H(s_i^{t_0}) = I(S^t | s_i^{t_0})$. In the other extreme case, the unit state $s_i^{t_0}$ is completely irrelevant to the system state $S^t$, the information is minimum at $I(S^t; s_i^{t_0}) = 0$.

The decay of this mutual information over time (as $t \to \infty$) is then a measure of the extent that the unit's state trajectory is affected by an instantaneous state of the unit. In other words, it measures the 'dynamical importance' of the unit. If the mutual information reaches zero quickly, then the state of the unit has a short-lasting effect on the collective behaviour of the system. The longer it takes for the mutual information to reach zero, the more influential is the unit to the system's behaviour. We call the time it takes for the mutual information to reach zero the IDT of a unit.

### 2.1.4. Defining the information dissipation time of a unit

At each time step, the information stored in a unit's state $s_i^t$ is partially transmitted to the next states of its nearest-neighbours [42,43], which, in turn, transmit it to their nearest-neighbours, and so on. The state of unit $s$ at time $t$ dictates the system state at the same time $t$ to the amount of

$$I_0^k \equiv I(S^t; s^t) = I(s^t; s^t) = H(s^t), \quad (2.2)$$

with the understanding that unit $s$ has $k$ interactions. We use the notation $I_0^k$ instead of $I_0^k$, because all units that have $k$ interactions are indistinguishable in our model. At time $t + 1$, the system state is still influenced by the unit's state $s^t$, the amount of which is given by

$$I_1^k = I(h^{t+1}; s^t). \quad (2.3)$$

As a result, a unit with $k$ connections locally dissipates its information at a ratio $I_1^k / I_0^k$ per time step. Here, we use the observation that the information about a unit's state $s^t$, which is at first present at the unit itself at the maximum amount

$H(s^t)$, can be only transferred at time $t + 1$ to the direct neighbours $h$ of $s$, through nearest-neighbour interactions.

At subsequent time steps ($t + 2$ and onward), the information about the unit with an amount of $I_1^k$ will dissipate further into the network at a constant average ratio

$$\hat{I} = \sum_m q(m) \cdot \frac{I_1^{m+1}}{I_0^{m+1}}. \quad (2.4)$$

from its neighbours, neighbours-of-neighbours, etc. This is due to the absence of degree–degree correlations or other structural bias in the network. That is, the distribution $q(m)$ of the degrees of a unit's neighbours (and neighbours-of-neighbours) does not depend on its own degree $k$. Here, $q(m) = (m + 1)p(m + 1)\langle m \rangle^{-1}$ is the probability distribution of the number of additional interactions that a nearest-neighbour unit contains besides the interaction with unit $s$, or the interaction with a neighbour of unit $s$, etc., called the excess degree distribution [44]. As a consequence, the dissemination of information of all nodes occurs at an equal ratio per time step except for the initial amount of information $I_1^k$, which the $k$ neighbour states contain at time $t + 1$, which depends on the degree $k$ of the unit. Note that this definition of $\hat{I}$ ignores the knowledge that the source node has exactly $k$ interactions, which at first glance may impact the ability of the neighbours to dissipate information. However, this simplification is self-consistent, namely we will show that $I_1^k$ diminishes for increasing $k$: this reduces the dissipation of information of its direct neighbours, which, in turn, reduces $I_1^k$ for increasing $k$, so that our conclusion that $I_1^k$ diminishes for increasing $k$ remains valid. See also appendix A for a second line of reasoning, about information flowing back to the unit $s$.

In general, the ratio per time step at which the information about $s_i^t$ dissipates from $t + 2$ and onward equals $\hat{I}$ up to an 'efficiency factor' that depends on the state–state correlations implied by the conditional transition probabilities $p(s_i^{t+1} | s_j^t)$. For example, if $s_A^t$ dictates 20% of the information stored in its neighbour state $s_B^{t+1}$, and $s_B^{t+1}$, in turn, dictates 10% of the information in $s_C^{t+2}$, then $I(s_A^t; s_C^{t+2})$ may not necessarily equal $20\% \times 10\% = 2\%$ of the information $H(s_C^{t+2})$ stored in $s_C^{t+2}$. That is, in one extreme, $s_B^{t+1}$ may use different state variables to influence $s_C^{t+2}$ than the variables that were influenced by $s_A^t$, in which case $I(s_A^t; s_C^{t+2})$ is zero, and the information transmission is inefficient. In the other extreme, if $s_B^{t+1}$ uses only state variables that were set by $s_A^t$ to influence $s_C^{t+2}$, then passing on $A$'s information is optimally efficient and $I(s_A^t; s_C^{t+2}) = 10\%$. Therefore, we assume that at every time step from time $t + 2$ onward, the ratio of information about a unit that is passed on is $c_{eff} \cdot \hat{I}$, i.e. corrected by a constant factor $0 \leq c_{eff} \leq 1/\hat{I}$ that depends on the similarity of dynamics of the units. It is non-trivial to calculate $c_{eff}$ but its bounds are sufficient for our proceeding.

Next, we can define the IDT of a unit. The number of time steps it takes for the information in the network about unit $s$ with degree $k$ to reach an arbitrarily small constant $\varepsilon$ is

$$D(s) = \log_{c_{eff} \hat{I}} \left[ \frac{\varepsilon}{I_1^k} \right] = \frac{\log \varepsilon - \log I_1^k}{\log c_{eff} + \log \hat{I}}. \quad (2.5)$$

Note that $D(s)$ is not equivalent to the classical correlation length. The correlation length is a measure of the time it takes for a unit to lose a certain *fraction* of its original correlation with the system state, instead of the time it takes for the unit to reach a certain *absolute value* of correlation. For our purpose of comparing the dynamical impact of units, the correlation length would not be a suitable measure. For example, if unit A has a large initial correlation with the system state and another unit B has a small initial correlation, but the halftime of their correlation is equal, then, in total, we consider A to have more impact on the system's state because it dictates more bits of information of the system state.

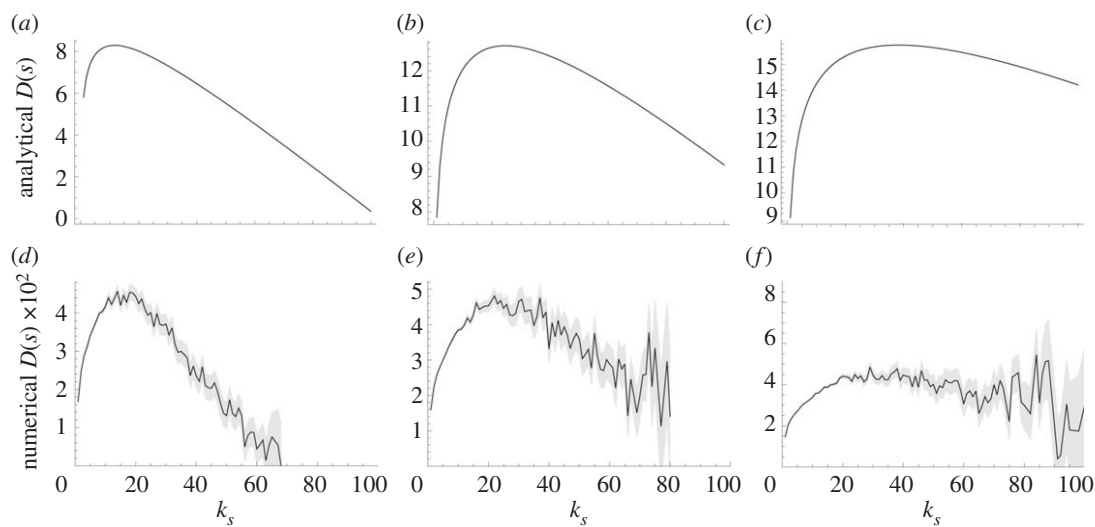

**Figure 1.** The dynamical impact $D(s)$ of a ferromagnetic Ising spin $s$ as function of its connectivity $k_s$, from evaluating the analytical $D(s)$ in equation (2.5) as well as from numerical experiments. For the analytical calculations, we used Glauber dynamics to describe the behaviour of the units; for the computer experiments, we used the Metropolis–Hastings algorithm. For the latter, we simulate a network of 6000 spins with a power-law degree distribution $p(k) \propto k^{-1.6}$; the plots are the mean of six realizations, each of which generated 90 000 time series of unit states that lead up to the same system state, which was chosen randomly after equilibration. The grey area is within two times the standard error of the mean IDT of a unit with a given connectivity. (a) $T = 2.0$, (b) $T = 2.5$, (c) $T = 2.75$, (d) $T = 9.0$, (e) $T = 12$ and (f) $T = 14$.

## 2.2. Diminishing information dissipation time of hubs

As a function of the degree $k$ of unit $s$, the unit's IDT satisfies

$$D(s) \propto \text{const} + \log I_1^k, \tag{2.6}$$

because $\hat{I}$, $c$ and $\varepsilon$ are independent of the unit's degree. Here, the proportionality factor equals $-(\log c_{\text{eff}} + \log \hat{I})^{-1}$, which is non-negative, because the dissipation ratio $c_{\text{eff}} \cdot \hat{I}$ is at most 1, and the additive constant equals $-\log \varepsilon$, which is positive as long as $\varepsilon < 1$. Because the logarithm preserves order, to show that the IDT diminishes for high-degree units, it is sufficient to show that $I_1^k$ decreases to a constant, as $k \to \infty$, which we do next.

The range of the quantity $I_1^k$ is

$$0 \le I_1^k \le \sum_{s_i^{t+1} \in h_i^{t+1}} I(s_j^{t+1}; s_i^t), \tag{2.7}$$

due to the conditional independence among the neighbour states $s_j^{t+1}$ given the node state $s_i^t$. In the average case, the upper bound can be written as $k \cdot \langle I(s_j^{t+1}; s_i^t) \rangle_{k_j}$, and we can write $I_1^k$ as

$$I_1^k = U(k) \cdot k \cdot T(k), \quad \text{where}$$
$$T(k) = \langle I(s_j^{t+1}; s_i^t) \rangle_{k_j}, \tag{2.8}$$

where $T(k)$ is the information in a neighbour unit's next state averaged over its degree, and $U(k)$ is the degree of 'uniqueness' of the next states of the neighbours. The operator $\langle \cdot \rangle_{k_j}$ denotes an average over the degree $k_j$ of a neighbour unit $s_j$, i.e. weighted by the excess degree distribution $q(k_j - 1)$. In one extreme, the uniqueness function $U(k)$ equals unity in case the information of a neighbour does not overlap with that of any other neighbour unit of $s_i^t$, i.e. the neighbour states do not correlate. It is less than unity to the extent that information does overlap between neighbour units, but is never negative. See §S3 in the electronic supplementary material for a detailed derivation of an exact expression and bounds of the uniqueness function $U(k)$.

Because the factor $U(k) \cdot k$ is at most a linear growing function of $k$, a sufficient condition for $D(s_i)$ to diminish as $k \to \infty$ is for $T(k)$ to decrease to zero more strongly than linear in $k$. After a few steps of algebra (see appendix), we find that

$$T(k+1) = \alpha \cdot T(k), \quad \text{where } \alpha \le 1. \tag{2.9}$$

Here, equality for $\alpha$ only holds in the degenerate case where only a single state is accessible to the units. In words, we find that the expected value of $T(k)$ converges downward to a constant at an exponential rate as $k \to \infty$. Because each term is multiplied by a factor $\alpha \le 1$, this convergence is downward for most systems but never upward even for degenerate system dynamics.

## 2.3. Numerical experiments with networks of Ising spins

For our experimental validation, we calculate the IDT $D(s)$ of 6000 ferromagnetic spins with nearest-neighbour interactions in a heavy-tailed network in numerical experiments and find that it, indeed, diminishes for highly connected spins. In figure 1, we show the numerical results and compare them with the analytical results, i.e. evaluating equation (2.5).

The analytical calculations use the single-site Glauber dynamics [45] to describe how each spin updates its state depending on the states of its neighbours. In this dynamics, at each time step, a single spin chooses its next state according to its stationary distribution of state, which would be induced if its nearest-neighbour spin states would be fixed to their instantaneous value (LTE). We calculate the upper bound of $D(s)$ by setting $U(k) = 1$, that is, all information about a unit's state is assumed to be unique that optimizes its IDT. A different constant value for $U(k)$ would merely scale the vertical axis.

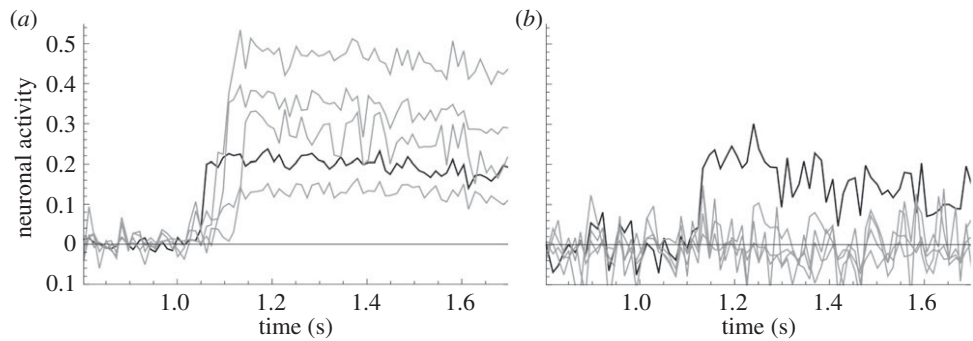

**Figure 2.** The level of activity of a set of neurons under a microscope as function of time, after seeding one neuron with an electrical potential (black line). The activity was measured by changes in calcium ion concentrations. These concentrations were detected by imaging fluorescence levels relative to the average fluorescence of the neurons (activity 0) measured prior to activation. In the sparse cultures with few synapses per neuron, the stimulated neuron evokes a network burst of activity in all other neurons in the field after a short delay. By contrast, in the dense cultures with many synapses per neuron, only the stimulated neuron has an increased potential. The data for these plots were kindly provided by Ivenshitz & Segal [41]. (a) Low connectivity and (b) high connectivity.

We perform computer simulations to produce time series of the states of 6000 ferromagnetic Ising spins and measure the dynamical importance of each unit by regression. For each temperature value, we generate six random networks with $p(k) \propto k^{-\gamma}$ for $\gamma = 1.6$ and record the state of each spin at 90 000 time steps. The state of each unit is updated using the Metropolis–Hastings algorithm instead of the Glauber update rule to show generality. In the Metropolis–Hastings algorithm, a spin will always flip its state if it lowers the interaction energy; higher energy states are chosen with a probability that decreases exponentially as function of the energy increase. Of the resulting time series of the unit states, we computed the time $d_i$ where $I(s_1^{t+d_i}, ...., s_N^{t+d_i}; s_i^t) = \varepsilon$ of each unit $s_i$ by regression. This is semantically equivalent to $D(s_i)$ but does not assume a locally tree-like structure or a uniform information dissipation rate $\hat{I}$. In addition, it ignores the problem of correlation (see appendix A). See section S1 in the electronic supplementary material for methodological details; see section S2 in the electronic supplementary material for results using higher values of the exponent $\gamma$. The results are presented in figure 1.

## 2.4. Empirical evidence

We present empirical measurements from the literature of the impact of units on the behaviour of three different systems, namely networks of neurons, social networks and protein dynamics. These systems are commonly modelled using a Gibbs measure to describe the unit dynamics. In each case, the highly connected units turn out to have a saturating or decreasing impact on the behaviour of the system. This provides qualitative evidence that our IDT, indeed, characterizes the dynamical importance of a unit, and, consequently, that highly connected units have a diminishing dynamical importance in a wide variety of complex systems. In each study, it remains an open question which mechanism is responsible for the observed phenomenon. Our work proposes a new candidate explanation for the underlying cause for each case, namely that it is an inherent property of the type of dynamics that govern the units.

The first evidence is found in the signal processing of *in vitro* networks of neurons [41]. The denser neurons are placed in a specially prepared Petri dish, the more connections (synapses) each neuron creates with other neurons. In their experiments, Ivenshitz and Segal found that sparsely connected neurons are capable of transmitting their electrical potential to neighbouring neurons, whereas densely connected neurons are unable to trigger network activity even if they are depolarized in order to discharge several action potentials. Their results are summarized in figure 2. In search for the underlying cause, the authors exclude some obvious candidates, such as the ratio of excitatory versus inhibitory connections, the presence of compounds that stimulate neuronal excitability and the size of individual postsynaptic responses. Although the authors do find tell–tale correlations, for example, between the network density and the structure of the dendritic trees, they conclude that the phenomenon is not yet understood. Note that in this experiment, the sparsely connected neuron is embedded in a sparsely connected neural network, whereas the densely connected neuron is in a dense network. A further validation would come from a densely connected neuron embedded in a sparse network in order to disentangle the network's contribution from the individual effect.

Second, in a person-to-person recommendation network consisting of four million persons, Leskovec *et al.* [39] found that the most active recommenders are not necessarily the most successful. In the setting of word-of-mouth marketing among friends in the social networks, the adoption rate of recommendations saturates or even diminishes for the highly active recommenders, which is shown in figure 3 for four product categories. This observation is remarkable, because in the dataset, the receiver of a recommendation does not know how many other persons receive it as well. As a possible explanation, the authors hypothesize that widely recommended products may not be suitable for viral marketing. Nevertheless, the underlying cause remains an open question. We propose an additional hypothesis, namely that highly active recommenders have a diminishing impact on the opinion forming of others in the social network. In fact, the model of Ising spins in our numerical experiments is a widely used model for opinion forming in social networks [14–16,18,20]. As a consequence, the results in figure 1 may be interpreted as estimating the dynamical impact of a person's opinion as function of the number of friends that he debates his opinion with.

The third empirical evidence is found in the evolutionary conservation of human proteins [40]. According to the neutral model of molecular evolution, most successful mutations in proteins are irrelevant to the functioning of the system of

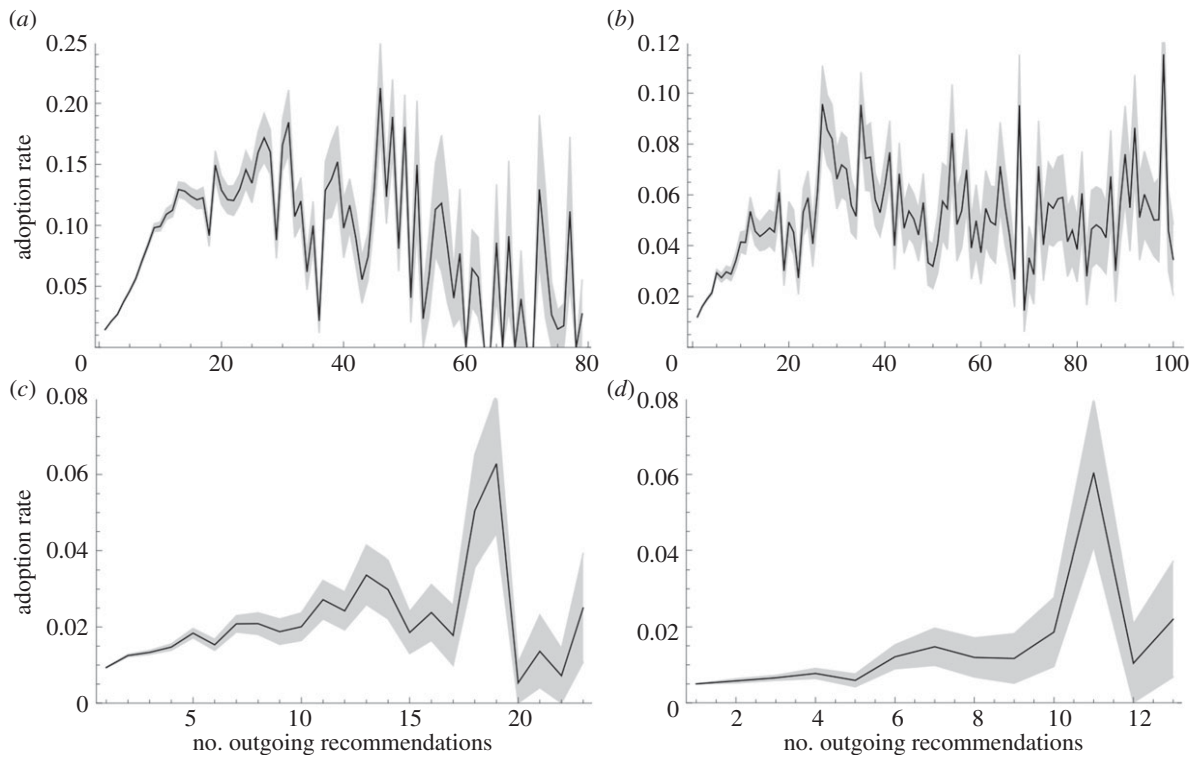

**Figure 3.** The success of a person's recommendation of a product as function of the number of recommendations that he sent. A person could recommend a product to friends only after he purchased the product himself. The success is measured as a normalized rate of receivers buying the product upon the recommendation. The normalization counts each product purchase equally in terms of the system's dynamics, as follows: if a person receives multiple recommendations for the same product from different senders, a 'successful purchase' is only accounted to one of the senders. The grey area is within 1 s.e.m. The total recommendation network consists of four million persons who made 16 million recommendations about half a million products. The subnetworks of the books and DVDs categories are by far the largest and most significant, with 73% of the persons recommending books and 52% of the recommendations concerning DVDs. The data for these plots were kindly provided by Leskovec *et al.* [39]. (a) DVD, (b) books, (c) music and (d) video.

protein–protein interactions [46]. This means that the evolutionary conservation of a protein is a measure of the intolerance of the organism to a mutation to that protein, i.e. it is a measure of the dynamical importance of the protein to the reproducibility of the organism [47]. Brown & Jurisica [40] measured the conservation of human proteins by mapping the human protein–protein interaction network to that of mice and rats using 'orthologues', which is shown in figure 4. Two proteins in different species are orthologous if they descend from a single protein of the last common ancestor. Their analysis reveals that the conservation of highly connected proteins is inversely related with their connectivity. Again, this is consistent with our analytical prediction. The authors conjecture that this effect may be due to the overall high conservation rate, approaching the maximum of 1 and therefore affecting the statistics. We suggest that it may indeed be an inherent property of protein interaction dynamics.

## 3. Discussion

We find that various research areas encounter a diminishing dynamical impact of hubs that is unexplained. Our analysis demonstrates that this phenomenon could be caused by the combination of unit dynamics and the topology of their interactions. We show that in large Markov networks, the dynamical behaviour of highly connected units have a low impact on the dynamical behaviour of the system as a whole, in the case where units choose their next state depending on the interaction potential induced by their nearest-neighbours.

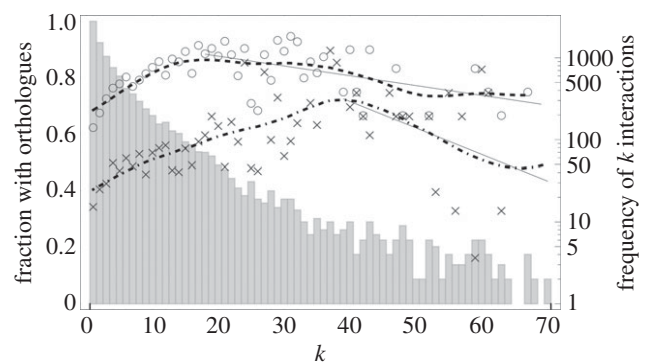

**Figure 4.** The fraction of evolutionary conservation of human proteins as a function of their connectivity $k$. The fraction of conservation is measured as the fraction of proteins that have an orthologous protein in the mouse (circles) and the rat (crosses). The dashed and dot-dashed curves show the trend of the conservation rates compared with mice and rates, respectively. They are calculated using a Gaussian smoothing kernel with a standard deviation of 10 data points. To evaluate the significance of the downward trend of both conservation rates, we performed a least-squares linear regression of the original data points starting from the peaks in the trend lines up to $k = 70$. For the fraction of orthologues with mice, the slope of the regression line is $-0.00347 \pm 0.00111$ (mean and standard error); with rats, the slope is $-0.00937 \pm 0.00594$. The vertical bars denote the number of proteins with $k$ interactions in the human protein–protein interaction network (logarithmic scale). The data for these plots were kindly provided by Brown & Jurisica [40].

For highly connected units, this type of dynamics enables the LTE assumption, originally used for describing radiative transport in a gas or plasma. To illustrate LTE, there is no single temperature value that characterizes an entire star: the outer shell is cooler than the core. Nonetheless, the mean free path of a moving photon inside a star is much smaller than the temperature gradient, so on a small timescale, the photon's movement can be approximated using a local temperature value. A similar effect is found in various systems of coupled units, such as social networks, gene regulatory networks and brain networks. In such systems, the internal dynamics of a unit is often faster than a change of the local interaction potential, leading to a multi-scale description. Intuitive examples are the social interactions in blog websites, discussion groups or product recommendation services. Here, changes that affect a person are relatively slow so that he can assimilate his internal state-of-mind (the unit's microstate) to his new local network of friendships and the set of personal messages he received, before he makes the decision to add a new friend or send a reply (the unit's macrostate). Indeed, this intuition combined with our analysis is consistent with multiple observations in social networks. Watts & Doods [48] numerically explored the importance of 'influentials', a minority of individuals who influence an exceptional number of their peers. They find counter to intuition that large cascades of influence are usually not driven by influentials, but rather by a critical mass of easily influenced individuals. Granovetter [49] found that even though hubs gather information from different parts of the social network and transmit it, the clustering and centrality of a node provide better characteristics for diffusing the innovator [50]. Rogers [51] found experimentally that the innovator is usually an individual in the periphery of the network, with few contacts with other individuals.

Our approach can be interpreted in the context of how dynamical systems intrinsically process information [42,43, 52–56]. That is, the state of each unit can be viewed as a (hidden) storage of information. As one unit interacts with another unit, part of its information is transferred to the state of the other unit (and vice versa). Over time, the information that was stored in the instantaneous state of one unit percolates through the interactions in the system, and at the same time it decays owing to thermal noise or randomness. The longer this information is retained in the system state, the more the unit's state determines the state trajectory of the system. This is a measure of the dynamical importance of the unit, which we quantify by $D(s)$.

Our work contributes to the understanding of the behaviour of complex systems at a conceptual level. Our results suggest that the concept of information processing can be used, as a general framework, to infer how dynamical units work together to produce the system's behaviour. The inputs to this inference are both the rules of unit dynamics as well as the topology of interactions, which contrasts with most complex systems research. A popular approach to infer the importance of units in general as topology-only measures such as connectedness and betweenness-centrality [28,30,57–62], following the intuition that well-connected or centrally located units must be important to the behaviour of the system. We demonstrate that this intuition is not necessarily true. A more realistic approach is to consider to simulate a simple process on the topology, such as the percolation of particles [63], magnetic spin interactions [3,6,14,20,37,64–72] or the synchronization of oscillators [37,60,73–80]. The dynamical importance of a unit in a such model is then translated to that of the complex system under investigation. Among the 'totalistic' approaches that consider the dynamics and interaction topology simultaneously, a common method to infer a unit's dynamical importance is to perform 'knock-out' experiments [29–31]. That is, experimentally removing or altering a unit and observing the difference in the system's behaviour. This is a measure of how robust the system is to a perturbation, however, and care must be taken to translate robustness into dynamical importance. In case the perturbation is not part of the natural behaviour of the system, then the perturbed system is not a representative model of the original system. To illustrate, we find that highly connected ferromagnetic spins hardly explain the observed dynamical behaviour of a system, even though removing such a spin would have a large impact on the average magnetization, stability and critical temperature [81,82]. In summary, our work is an important step towards a unified framework for understanding the interplay of the unit dynamics and network topology from which the system's behaviour emerges.

Acknowledgements. We thank Carlos P. Fitzsimons for helping us find and interpret empirical evidence from the field of neurobiology. We also thank Gregor Chliamovitch and Omri Har-Shemesh for their feedback on the mathematical derivations.

Funding statement. We acknowledge the financial support of the Future and Emerging Technologies (FET) programme within the Seventh Framework Programme (FP7) for Research of the European Commission, under the FET-Proactive grant agreement TOPDRIM, number FP7-ICT-318121, as well as under the FET-Proactive grant agreement Sophocles, number FP7-ICT-317534. P.M.A.S. acknowledges the NTU Complexity Programme in Singapore and the Leading Scientist Programme of the Government of the Russian Federation, under contract no .11.G34.31.0019.

# Appendix A

## A.1. Limiting behaviour of $p(s_i^{t+1} = q)$ as $k \longrightarrow \infty$

Using equation (2.1), the prior probability of a unit's state can be written as

$$p(s_i^{t+1} = q) = \sum_{r = \{r_1, \dots, r_k\} \in \Sigma^k} p(h_i^t = r) \cdot e^{-\sum_{j=1}^{k} e(q, r_j)/T} \cdot Z_k^{-1}, \quad \text{(A 1)}$$

where $Z_k$ is the partition function for a unit with $k$ edges. As $k \gg |\Sigma|$, the set of interaction energies starts to follow a stationary distribution of nearest-neighbour states, and the expression can be approximated as

$$p(s_i^t = q) = e^{-k\langle e_q \rangle/T} \cdot Z_k^{-1}. \quad \text{(A 2)}$$

Here, $\langle e_q \rangle$ is the expected interaction energy of the state $q$ with one neighbour, averaged over the neighbours' state distribution. If an edge is added to such a unit, the expression becomes (the subscript $k + 1$ denotes the degree of the node as a reminder)

$$p_{k+1}(s_i^t = q) = e^{-(k+1)\langle e_q \rangle/T} \cdot Z_{k+1}^{-1}$$
$$= e^{-\langle e_q \rangle/T} \cdot e^{-k\langle e_q \rangle/T} \cdot Z_{k+1}^{-1}. \quad \text{(A 3)}$$

In words, the energy term for each state $q$ is multiplied by a factor $e^{-\langle e_q \rangle/T}$ that depends on the state but is constant with respect to $k$. (The partition function changes with $k$ to suitably normalize the new terms, but it does not depend on $q$ and so is not responsible for moving probability mass.) That is, as $k$ grows, the probability of the state $q$ with the lowest expected interaction energy approaches unity; the probabilities of all other states will approach zero. The approaches are exponential, because the multiplying factors do not depend on $k$. If there are $m$ states with the lowest interaction energies (multiplicity $m$), then each probability of these states will approach $1/m$.

## A.2. Deriving an upper bound on $\alpha$ in $T(k+1) = \alpha \cdot T(k)$

First, we write $T(k)$ as an expected mutual information between the state of a unit and the next state of its neighbour, where the average is taken over the degree of the neighbour unit:

$$T(k) = \langle H(s_i^t) - H(s_i^t|s_j^{t+1}) \rangle_{k_j}. \qquad (A\,4)$$

We will now study how $T(k)$ behaves as $k$ grows for large $k$. By definition, both entropy terms are non-negative, and $H(s_i^t|s_j^{t+1}) \leq H(s_i^t)$. In §A.1 of this appendix, we find that the prior probabilities of the state of a high-degree unit exponentially approach either zero from above or a constant from below. In the following, we assume that this constant is unity for the sake of simplicity, i.e. that there is only one state with the lowest possible interaction energy.

$$
\begin{aligned}
H(s_i^t) &= -\sum_{q\in\Sigma} p(s_i^t = q) \log p(s_i^t = q), \\
&= -\sum_{q\in\Sigma^+} (1 - b_q^{-k}) \log(1 - b_q^{-k}) - \sum_{q\in\Sigma^-} b_q^{-k} \log b_q^{-k}, \\
&= -\sum_{q\in\Sigma^+} (1 - b_q^{-k}) \log(1 - b_q^{-k}) + k \sum_{q\in\Sigma^-} b_q^{-k} \log b_q, \\
&\approx k \sum_{q\in\Sigma^-} b_q^{-k} \log b_q, \\
&= O(k \cdot x^{-k}).
\end{aligned}
$$
$$(A\,5)$$

In words, the first entropy term eventually goes to zero exponentially as function of the degree of a unit. Because this entropy term is the upper bound on the function $T(k)$, there are three possibilities for the behaviour of $T(k)$. The first option is that $T(k)$ is zero for all $k$, which is a degenerate system without dynamical behaviour. The second option is that $T(k)$ is a monotonically decreasing function of $k$, and the third option is that $T(k)$ first increases and then decreases as function of $k$. In both cases, for large $k$ the function, $T(k)$ must approach zero exponentially.

In summary, we find that for large $k$

$$T(k+1) = \alpha \cdot T(k), \text{ where } \alpha < 1. \qquad (A\,6)$$

The assumption of multiplicity unity of the lowest interaction energy is not essential. If this assumption is relieved, then in step 3 of equation (A 5), then the first term does not become zero but a positive constant. It may be possible that a system where $T(k)$ equals this constant across $k$ is not degenerate, in contrast to the case of multiplicity unity, so in this case, we must relax the condition in equation (A 6) to include the possibility that all units are equally important, i.e. $\alpha \leq 1$. This still makes it impossible for the impact of a unit to keep increasing as its degree grows.

## A.3. Information flowing back to a high-degree unit

In the main text, we simplify the information flow through the network by assuming that the information at the amount $I_i^k$ stored in the neighbours of a unit flows onward into the network, and does not flow back to the unit. Here, we rationalize that this assumption is appropriate for high-degree units.

Suppose that at time $t + 1$, the neighbour unit $s_j$ stores $I(s_i^t; s_j^{t+1})$ bits of information about the state $s_i^t$. At time $t + 2$, part of this information will be stored by two variables: the unit's own state $s_i^{t+2}$ and the combined variable of neighbour-of-neighbour states $\{s_{j1}, ..., s_{jk_j}\}$. In order for the IDT $D(s_i)$ of unit $s_i$ to be affected by the information that flows back, this information must add a (significant) amount to the total information at time $t + 2$. We argue however that this amount is insignificant, i.e.

$$
\begin{aligned}
&I(s_i^t; S^{t+2}) - I(s_i^t; \{s_{j1}^{t+2}, ..., s_{jk_j}^{t+2}\}) \\
&= I(s_i^t; s_i^{t+2}|\{s_{j1}^{t+2}, ..., s_{jk_j}^{t+2}\}) \xrightarrow{k_j \to \infty} 0.
\end{aligned}
$$
$$(A\,7)$$

The term $I(s_i^t; s_i^{t+2}|\{s_{j1}^{t+2}, ..., s_{jk_j}^{t+2}\})$ is the conditional mutual information. Intuitively, it is the information that $s_i^{t+2}$ stores about $s_i^t$ which is not already present in the states $\{s_{j1}^{t+2}, ..., s_{jk_j}^{t+2}\}$.

The maximum amount of information that a variable can store about other variables is its entropy, by definition. It follows from sections A.1 and A.2 of appendix that the entropy of a high-degree unit is lower than the average entropy of a unit. In fact, in the case of multiplicity unity of the lowest interaction energy the capacity of a unit goes to zero as $k \to \infty$. For this case, this proves that $I(s_i^t; s_i^{t+2}|\{s_{j1}^{t+2}, ..., s_{jk_j}^{t+2}\})$, indeed, goes to zero. For higher multiplicities, we observe that the entropy $H(s_i^{t+2})$ is still (much) smaller than the total entropy of the neighbours of a neighbour $H(s_{j1}^{t+2}) + H(s_{j2}^{t+2}|s_{j1}^{t+2}) + \cdots$. Therefore, the information $I(s_i^t; s_i^{t+2})$ that flows back is (much) smaller than $I(s_i^t; \{s_{j1}^{t+2}, ..., s_{jk_j}^{t+2}\})$, and the conditional variant is presumably smaller still. Therefore, we assume that also in this case, the information that flows back has an insignificant effect on $D(s_i)$.

## A.4. A note on causation versus correlation

In the general case, the mutual information $I(s_x^t; s_y^{t_0})$ between the state of unit $s_x$ at time $t_0$ and another unit's state $s_y$ at time $t$ is the sum of two parts: $I_{\text{causal}}$, which is information that is due to a causal relation between the state variables, and $I_{\text{corr}}$, which is information due to 'correlation' that does not overlap with the causal information. Correlation occurs if the units $s_x$ and $s_y$ both causally depend on a third 'external' variable $e$ in a similar manner, i.e. such that $I(e; (s_x^t, s_y^{t_0})^{\mathsf{T}}) < I(e; s_x^t) + I(e; s_y^{t_0})$. This can lead to a non-zero mutual information $I(s_x^t; s_y^{t_0})$ among these two units, even if the two units would not directly depend on each other in a causal manner [83,84].

For this reason, we do not directly calculate the dependence of $I(S^t; s^{t_0})$ on the time variable $t$ in order to calculate the IDT of a unit $s$. It would be difficult to tell how much of this information is non-causal at every time point. In order to find this out, we would have to understand exactly how each bit of information is passed onward through the system, from one state variable to the next, which we do not yet understand at this time.

To prevent measuring the non-causal information present in the network, we use local single-step 'kernels' of information diffusion, namely the $I_1^k/I_0^k$ as discussed previously. The information $I_0^k$ is trivially of causal nature (i.e. non-causal information is zero), because it is fully stored in the state of the unit itself. Although, in the general case, $I_1^k$ may consist of a significant non-causal part, in our model, we assume it to be zero or at most an insignificant amount. The rationale is that units do not self-interact (no self-loops), and the network is locally tree-like: if $s_x$ and $s_y$ are direct neighbours, then there is no third $s_z$ with 'short' interaction pathways to both $s_x$ and $s_y$. The only way that non-causal (i.e. not due to $s_x^t$ influencing $s_y^{t+1}$) information can be created between $s_x^t$ and $s_y^{t+1}$ is through the pair of interaction paths $s_z^{t'} \to \cdots \to s_y^{t-1} \to s_x^t$ and $s_z^{t'} \to \cdots \to s_y^{t+1}$, where $t' < t - 1$. That is, one and the same state variable $s_z^{t'}$ must causally influence both $s_x^t$ and $s_y^{t+1}$, where it can reach $s_x$ only through $s_y$. We expect any thusly induced non-causal information in $I(s_y^{t+1}; s_x^t)$ is insignificant compared with the causal information through $s_x^t \to s_y^{t+1}$, and the reason is threefold. First, the minimum lengths of the two interaction paths from $s_z$ are two and three interactions, respectively, where information is lost through each interaction due to its stochastic nature. Second, of the information that remains, not all information $I(s_z^{t'}; s_x^t)$ may overlap with $I(s_z^{t'}; s_y^{t+1})$, but even if it does, then the 'correlation part' of the mutual information $I(s_y^{t+1}; s_x^t)$ due to this overlap is upper bounded by their minimum: $\min \{ I(s_z^{t'}; s_x^t), I(s_z^{t'}; s_y^{t+1}) \}$. Third, the mutual information due to correlation may, in general, overlap with the causal information, i.e. both pieces of information may be partly about the same state variables. That is, the $I_{corr}$ part of $I(s_y^{t+1}; s_x^t)$, which is the error of our assumption, is only that part of the information-due-to-correlation that is not explained by (contained in) $I_{causal}$. The final step is the observation that $I_1^k$ is the combination of all $I(s_y^{t+1}; s_x^t)$ for all neighbour units $s_y \in h_x$.